\documentclass[times, 10pt,twocolumn]{article} 
\usepackage{latex8}
\usepackage{times}
\usepackage{graphicx}
\usepackage{float}
\usepackage{array}
\usepackage{booktabs}

\usepackage{amssymb}
\usepackage{pifont}
\newcommand{\xmark}{\ding{55}}%
\usepackage[font=small,labelfont=bf]{caption}
\usepackage{enumitem}
\usepackage{color}

\begin{document}

\title{TARN: A SDN-based Traffic Analysis Resistant Network Architecture}

\author{Lu Yu, Qing Wang, Geddings Barrineau, Jon Oakley, \\ 
Richard R. Brooks, and Kuang-Ching Wang\\
Department of Electrical and Computer Engineering\\
Clemson University, Clemson, SC, USA\\ 
$\lbrace$lyu,qw,cbarrin,joakley,rrb,kwang$\rbrace$@g.clemson.edu \\
}

\maketitle
\thispagestyle{empty}

\begin{abstract}
\noindent Destination IP prefix-based routing protocols are core to
Internet routing today.  Internet autonomous systems (AS) possess
fixed IP prefixes, while packets carry the intended destination AS's
prefix in their headers, in clear text.  As a result, network
communications can be easily identified using IP addresses and become
targets of a wide variety of attacks, such as DNS/IP filtering,
distributed Denial-of-Service (DDoS) attacks, man-in-the-middle (MITM) attacks, etc.  In this work, we explore an alternative network
architecture that fundamentally removes such vulnerabilities by
disassociating the relationship between IP prefixes and destination
networks, and by allowing any end-to-end communication session to have
dynamic, short-lived, and pseudo-random IP addresses drawn from a
range of IP prefixes rather than one.  The concept is seemingly
impossible to realize in today’s Internet. We demonstrate how this is
doable today with three different strategies using software defined
networking (SDN), and how this can be done at scale to transform the
Internet addressing and routing paradigms with the novel concept of a
distributed software defined Internet exchange (SDX).  The solution
works with both IPv4 and IPv6, whereas the latter provides higher
degrees of IP addressing freedom.  Prototypes based on Open vSwitches
(OVS) have been implemented for experimentation across the PEERING BGP
testbed.  The SDX solution not only provides a technically sustainable
pathway towards large-scale traffic analysis resistant network (TARN)
support, it also unveils a new business model for customer-driven,
customizable and trustable end-to-end network services.
\end{abstract}

%
\section{Introduction}
Internet was not designed with security and privacy in mind and is
littered with security holes.  The notorious network censorship, BGP
hijacking~\cite{nuclear,ice}, man-in-the-middle (MITM) attacks,
and distributed Denial-of-Service (DDoS) attacks~\cite{ozcelik2013detection,zhong2017ETCI}, to name a few, pose real threat to network security and user privacy.  These attacks all
rely on a seemingly innocuous and crucial element of today's Internet
Protocol (IP) to identify their victims -- the IP addresses of the
communicating source and destination.  At domain-level routing,
network operators rely on IP prefix-based BGP protocol for
inter-domain routing.  And what's worse, IP addresses carried in
packet headers are always in clear text.

Consequently, national firewalls easily restrict inhabitants' access to information by simply blocking the IP addresses of those websites~\cite{freedomhouse}. While encryption and virtual private networks (VPN) provide some protection, some countries use traffic analysis to further track citizens that access opposition web sites.  Similarly, victims of DDoS attacks and MITM attacks have nowhere to hide due to the static binding between the hosts and their uniquely identified IP addresses.

The proposed work explores an alternative end-to-end network
architecture called the Traffic Analysis Resistant Network (TARN) that
fundamentally removes the vulnerability due to fixed IP addresses of
communication sessions by using, instead, multiple short-lived,
seemingly random IP addresses.  By breaking the static binding between
hosts and IP addresses, TARN makes it impossible for attackers to
detect communication sessions by observing the source and destination
addresses.  While in theory traffic analysis methods can analyze other
properties of network traffic, e.g., protocol types~\cite{beasley2014survey}, packet sizes and
inter-packet timing~\cite{yu2015side}, to infer the nature of communication, IP
addresses are undeniably the prime references used by attackers today.  We recognize TARN induces
disruptive changes to multiple aspects of the Internet, ranging from
domain name lookup to packet forwarding; an incremental approach to
gradually introduce TARN into Internet is hence critical.  In this
paper, we give an overview of three implementation strategies that
realizes TARN at different scopes and discuss their implications.

TARN would not have been practical without two recent technology
trends -- software defined networking (SDN) and cloud-based virtual
network functions (VNF). With SDN, end-to-end communication sessions
can be identified and modified (in TARN's case, the IP addresses) with
fine grain packet header fields at flexible network locations.  With
cloud-based VNF, TARN functions can be scalably instantiated driven by
user and application demands. As discussed in section~\ref{sec:tarn},
besides the option of running TARN at an end host, a cloud-based
implementation of TARN, which we call a software defined network
exchange (SDX), can be operated by a provider much like the Internet
exchange providers today with advanced, highly customizable services
for communication sessions. We envision SDXs to offer end users a
logically centralized portal to request and configure services with
user-chosen attributes, with the actual TARN packet handling to be
executed across a distributed network of SDX sites.

The remainder of this paper is organized as follows.
Section~\ref{sec:rw} provides related work in the area of anonymity
network and a brief introduction of SDX.  In section~\ref{sec:tarn},
we introduce the TARN architecture and the implementation strategies.
Early experimentation results are discussed in section~\ref{sec:exp}. 
The paper concludes in section~\ref{sec:con} with potential challenges
and the future work.

%
\section{Related Work}\label{sec:rw}
Various proxy services such as The Onion Routing (Tor)~\cite{tor},
Psiphon~\cite{psiphon}, and I2P~\cite{i2p}, make traffic analysis
difficult by inserting intermediate nodes between packet source and
destination. The extent to which they can protect communication rely
on such intermediate nodes' ability to remain unknown to
adversaries. With their IP addresses being fixed, their success is
limited to the time it takes for adversaries to detect them and start
analyzing or blocking communications going through them. 

Researchers have found the increased latency due to the use of intermediate nodes in such solutions to be a major impediment to user acceptance~\cite{roberts2007circumvention}.  I2P is not very widely used.  Tor is commonly attacked by countries finding Tor entry nodes and blocking their IP addresses.  This can include active probing of
nodes suspected of providing entry to the Tor
network~\cite{chao2016china}.  Tor is mostly blocked by Chinese
government.  One Tor initiative avoids traffic blocking by using
``pluggable transports'' to obfuscate the traffic pattern for the
first hop of the proxy.  The issue with the traffic obfuscation
techniques is that they only obfuscate communication patterns, but not
the end hosts~\cite{houmansadr2013want}.  Another concern with proxy
networks is that users must trust the intermediate nodes, any of which
could be compromised to perform a MITM attack.

Decoy routing, a new alternative, inserts steganography information
into the header of SSL packets which is recognized by ``decoy
routers'' in the backbone, who re-route packets to their intended
destination~\cite{karlin2011decoy}.  One drawback is that nodes
currently have to attempt connections at random until they find the
decoy routing node.

There are other recent SDN-based solutions, e.g., the OpenFlow-based
random host mutation (OFRHM)~\cite{kampanakis2014sdn}, aiming to use
OpenFlow to randomly mutate source host IP addresses to avoid unwanted
traffic analysis.  However, OFRHM uses random addresses within the
same subnet of a source node.  

The notion of SDX was first described in 2013 by Feamster, et
al.~\cite{gupta2015sdx} as a Software Defined Internet Exchange,
applying SDN at Internet exchange points (IXP) to enable finer grain,
application specific peering beyond what BGP is capable of today.
Since then, various forms and applications of SDX have been
proposed. In~\cite{mambretti2014software}, SDX was defined as a
software defined networking exchange that supports interconnect of
multiple network domains at flexible layers (not limited to layer 3 or
BGP) via signaling among federated network controllers.  At the 2013
SDN Program Review workshop
~\cite{sdn2013report}, one more
form of SDX was discussed to be an exchange point for software defined
infrastructure (SDI) facilities, providing both SDN interconnect and
inline compute services that can be custom programmed via software
APIs.

\section{The TARN Architecture}
\label{sec:tarn}
\begin{figure}[!h]
\centering
\includegraphics[width=0.5\textwidth]{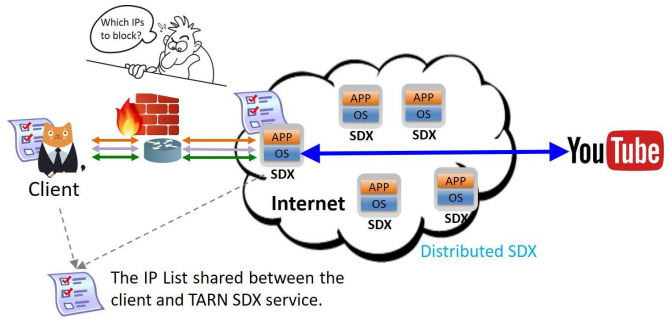}
\caption{The architecture of TARN for the use case of network censorship circumvention. A client wants to connect to YouTube that is blocked by the client's local ISP.}
\label{fig:tarn}
\end{figure}

\subsection{Overview}
\label{sub:overview}
The core concept of TARN is to bind the two ends of a communication
with dynamic, ephemeral and perpetually changing IP addresses.
Only authorized users who know the functions used for finding the
address ``hopping'' pattern can track the IP changes over time.  This
is similar to frequency-hopping spread spectrum
(FHSS)~\cite{schwartz2001frequency} for wireless communications.  In
this context, we are proposing IP hopping in address space.  By doing
so, TARN effectively hides the client or/and server from being
detected by malicious attackers.

To facilitate an introduction, we consider a use case of TARN --
network censorship circumvention.  The corresponding TARN architecture
is illustrated in Figure~\ref{fig:tarn}, in which the client wants to
access YouTube which is blocked by the client's local Internet service
provider (ISP).  The objective is to enable the client to access
YouTube without the ISP detecting and blocking the communication
between them.  We assume the ISP can monitor all traffic leaving and
entering the AS.

In this particular scenario, it is impractical to assume a client
would use randomized IP addresses.  Most ISPs control the assignment of a
fixed IP address to a client.  Only the server side (i.e., YouTube)
adopts IP hopping in this case.  As illustrated in
Figure~\ref{fig:tarn}, YouTube employs distributed SDX TARN service
across the Internet.  With the service, YouTube can customize its IP
hopping pattern and communicate it with the client as an authorized
customer.  The SDX provider would provide YouTube a number of
different IP prefixes where IP addresses can be flexibly and
dynamically used by the YouTube server over time. In this case, the
SDX provider will own sufficiently many IP prefixes that can be
allocated, most possibly through a lease, to a service provider as
YouTube.  Note that TARN does not suffer from IP blocking as most other privacy-preserving technologies, including Tor.

\subsection{Security features of TARN}
The fixed binding between hosts and IP addresses are crucial to today's Internet routing.  However, such scheme has also been taken advantage of by malicious attackers.  It enables IP/DNS filtering to become a common method used by nation-state firewalls to block opposition websites.  Network users who struggle against censorship really have only two options: VPN and proxy, neither of which can avoid being blocked once being identified.  Moreover, fixed IP addresses are also the prime references used by attackers to target their victims, which is a necessary step before staging any malicious attacks.  

As a result, breaking the rule of fixed IP addresses becomes a natural solution to thwart network attacks.  As the most innovative element of TARN, the idea of IP hopping is similar, but in many aspects superior to the \textit{fast flux} technique used by the notorious botnets to elude detection~\cite{fu2017stealthy}.  And the effectiveness of fast flux as an anti-detection measure has been thoroughly tested and validated in practice by the protracted game between botnets and law enforcement~\cite{fu2015analysis}.  Therefore, we envision that the adoption of TARN will mitigate a significant portion of network attacks.

\subsubsection{Massaging the IP hopping patterns}
We are fully aware that abnormally frequent change of IP addresses may be exploited by adversaries to identify TARN users.  We must account for the possible side-channel analysis on the IP hopping frequency of TARN users.  In this work, we ``massage'' the IP hopping pattern of TARN users' traffic to make it statistically identical to that of the regular network users~\cite{zhong2015stealthy}.  This allows traffic generated by TARN users to be seamlessly blended into the background traffic.  

The idea is to construct a hidden Markov model (HMM) that represents the IP hopping pattern of regular users and then use the model to determine the dwell time of each IP address for TARN users.  The inferred model is known as a deterministic HMM~\cite{yu2013inferring,yu2012stochastic,lu2013normalized}.  Unlike the standard HMM that has two sets of random processes: states and observations.  In a deterministic HMM, there is only a single set of random processes (time intervals between two hops in this case).  So the state transition labels are uniquely associated with an output alphabet.  

Given the inferred model, TARN service starts the timing side-channel massage by randomly selecting a start state in the model.  To determine the dwell time of the current address, a transition is taken from the current state and the corresponding time interval is waited before switching to the next address.  If there are more than one possible transitions out of current state, the transition is chosen randomly, weighed on the probability of each transition.  

Deep packet inspection (DPI) techniques censor traffic based on the application-layer content and can be used to identify TARN users.  We have developed pluggable transports for Tor that mimic protocols like NTP, video games, and smart grid sensors~\cite{zhong2015stealthy}.  Our pluggable transports encrypt a session and then translate the AES encrypted data into the syntax of another host protocol.  Integration of our pluggable transports into TARN for DPI circumvention is beyond the scope of this paper. 


\subsection{Implementation Strategies}
\label{sec:implementation}
The TARN architecture can assume multiple forms, from a
``host-only'' end-to-end solution to a distributed SDX solution.  We
discuss three different implementation strategies that serve different
needs and scopes. All designs are compatible with existing network
infrastructure and routing protocols.  While implementations vary,
they all adopt SDN for programmatic packet switching based on packet
flow attributes at different protocol layers (layer 2--layer 4).

Recall that the two core functions of TARN are $1)$ to unbind IP
prefixes from ASes; and $2)$ to allow sustainable communication
sessions between two ends with short-lived and seemingly
randomized IP addresses (not constrained to a single prefix).  To achieve
this, three design patterns are possible: end-host TARN agent, campus TARN gateway, and SDX TARN service, each of which assumes adifferent threat model.  Table~\ref{tab:synopsis} provides a synopsis of the tradeoffs for them.  Each design pattern implements IP rewriting where appropriate
to the specific threat model, i.e., the trust boundary.

\subsubsection{End-host TARN agent}
In the first pattern, a software agent sits on each end host, together
with a virtual SDN switch such as Open vSwitch
(OVS)~\cite{openvswitch}, to rewrite the destination and/or source IP
addresses before the packets leave the host.  The pattern assumes any
part of the network between two end hosts are potentially vulnerable
to traffic analysis attacks, and thus, requires address
rewrite to protect both ends' identity and accessibility.  This
design pattern typically suffices in allowing a person/client to
access a website, e.g., google.com, that would otherwise be blocked by
the person's ISP.
\begin{table}[!h]
\centering 
{\renewcommand{\arraystretch}{1}
\begin{tabular}{>{\centering}m{1.5cm} >{\centering}m{1.5cm}  >{\centering}m{1.5cm}  >{\centering\arraybackslash}m{2cm}} 
\toprule
\footnotesize\textbf{Design Pattern} & \footnotesize\textbf{Host application required} & \footnotesize\textbf{BGP route announcements} &\footnotesize\textbf{Trust boundary} \\ 
\hline 
\footnotesize
End-host TARN agent & \footnotesize\checkmark & \footnotesize\checkmark & \footnotesize Host-only \\ 
\hline
\footnotesize Campus TARN gateway & \footnotesize\xmark & \checkmark &\footnotesize Campus network \\
\hline
\footnotesize SDX-based TARN service & \footnotesize\xmark &\footnotesize\xmark & \footnotesize Campus network + Path to SDX facilities \\
\hline 
\end{tabular}}
\caption{Considerations for three TARN design patterns.} 
\label{tab:synopsis} 
\end{table}

The TARN agent running on the client-host will rewrite the destination
address of the outgoing requests to the website.  On the website
server’s end, either a TARN agent or a TARN gateway (see
section~\ref{subsec:gateway}) would be used to accept the incoming
requests and rewrite them into the correct, original web server
addresses.  One critical requirement to note -- the server side needs
to announce via BGP new prefixes that are not the organization's known
prefixes. Such prefixes can be scavenged from unused prefixes (a lot
more in IPv6 than in IPv4) in the near term.  In the long term, we
believe SDX’s can acquire large pools of IP prefixes for its
customers' scheduled reuse.  Such prefixes will only be held by any
organization for a relatively short time to prevent adversaries from
detecting their relationship.  The prefixes will only be made known to
registered customers through authenticated means to avoid adversaries
to fish for the addressing scheme.

\subsubsection{Campus TARN gateway}\label{subsec:gateway}
The second pattern relieves the end hosts' burden and runs the IP
rewrite actions at a campus gateway router, called the TARN gateway.
End hosts choosing to use the service set the TARN gateway their
default router.  This pattern assumes the entire campus network as a
trusted domain where TARN protection is not necessary.  This pattern
makes more sense to be adopted by major, multi-campus corporates that
need to have discrete anonymized inter-campus communications.  As a
result, the TARN gateways are expected to be deployed at all
participating campuses and their functions will be symmetric for any
inter-campus communication sessions.

\subsubsection{SDX TARN service}
As we explore different implementation strategies for TARN, the
concept of a distributed SDX surfaces as an attractive option.  Recent
studies have explored SDX in different forms and for different
purposes~\cite{gupta2015sdx}.

Thus far, a SDX has been considered as a SDN-empowered IXP facility -
a place where networks of multiple ASes interconnect and forward
packet traffic.  Today, there are hundreds of IXPs worldwide, and the
large IXPs can be interconnecting up to several hundred ASes and
forwarding a quarter billion IP addresses each
week~\cite{chatzis2013benefits}.  In our study, we consider a
distributed SDX with the following properties:

\begin{itemize}[itemsep=0mm]
\item Consisting of multiple distributed facilities spread across Internet;
\item Connecting ASes at flexible protocol layers, e.g., as layer-2 circuits or layer-3 routes;
\item Supporting custom \textit{packet transformation} in addition to \textit{packet switching}, programmable across any affiliated facilities; and
\item Serving each AS as a customer, and facilitating programmable customer-to-customer packet exchange (i.e., transformation and switching).
\end{itemize}

\begin{figure}[!h]
\centering
\includegraphics[width=0.45\textwidth]{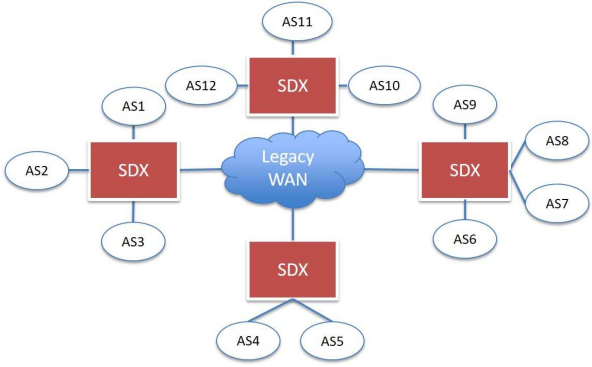}
\caption{Distributed SDX, where each participant AS connects of a SDX.}
\label{fig:distributedSDX}
\end{figure}

To achieve these properties, we expect each SDX facility to have a
SDN-based switching fabric for interconnecting customer ASes and
exchanging traffic between them, and a cloud-like compute
infrastructure that executes network functions in between any ASes for
custom packet transformation demands.  A SDX provider is expected to
operate sufficiently many facilities at strategic locations across
Internet.  Customers for such a SDX service will, most likely through
a provider portal, provision a range of network functions at any or
all of the provider's facilities to handle incoming and outgoing
traffic flows among theirs and others’ ASes.
Figure~\ref{fig:distributedSDX} illustrates the distributed SDX.
Gupta et. al have shown in~\cite{gupta2015sdx} that integrating SDX
with existing IXP infrastructure and conventional BGP-speaking ASes is
straightforward.  And our TARN service makes a compelling wide-area
traffic delivery application of SDX.

The SDX pattern consolidates all TARN packet transformation and
switching functions to a SDX facility, so the customers are primarily
networks, groups of networks or ASes (AS 43515, which is YouTube’s AS
number) rather than individual end users.  In the case of censorship
circumvention, the SDX TARN service appeals to customers like Google,
Youtube, NYT, Facebook, Twitter, etc, which are blocked by many
nation-state firewalls.  From each customer network, be it a campus
network or a CDN, we assume a secure channel between the customer and
the SDX facility.  For example, a layer-2 connection such as the
Internet2 Advanced Layer 2 Service (AL2S)~\cite{internet2} can be
used. In this case, each customer will have a dedicated landing place
which most possibly can be a virtual machine (VM), on which the TARN
configurations can be controlled completely by the end-user
organization.  As a result, the SDX facility can be viewed as a public
cloud, with many VMs (at least one per customer) interconnected by the
SDN fabric.

\subsection{Challenges}
\label{subsec:challenge}
We fully recognize that this transformative paradigm introduces
significant changes to multiple aspects of the Internet, ranging from
routing to system scalability.  Since TARN builds on existing Internet
infrastructure, to realize TARN in practice, we must identify these
challenges and provide sound solutions.

\subsubsection{IP address collision avoidance.} 
TARN works with both IPv4 and IPv6 networks.  Address collisions can
occur, while the vast IPv6 address space makes this mathematically impossible.
Therefore, the approach is particularly flexible with IPv6.  The
probability of having at least one collision when randomly mapping the
entire IPv4 space into IPv6 space is extremely small, approximately
$3.906\times 10^{-28}$ compared with the probability of being hit by a
meteor being $5.49\times 10^{-15}$.

As discussed later, with SDX, the impact of address collisions can be
minimized.  Without SDX, collisions will have to be resolved with
known duplicate address detection (DAD) methods and packet
re-transmissions.

\subsubsection{Efficient routing to/between dynamic addresses:} 
Traffic to these ``ephemeral'' addresses are routed to the desired
destination using BGP route injection, intermediate software defined
Internet exchanges (SDX), or a combination of the two.  We inject BGP
routes for these random and dynamic addresses beforehand and withdraw
them after use.  In today's Internet, BGP route updates converge
globally in just a few minutes.  And the propagation of newly
announced prefixes happens almost instantaneously.  We are fully aware
that ISPs today will not accept frequent BGP announcement, although
such operation is technically feasible and compatible with existing
Internet infrastructure.  The purpose of this work is to demonstrate
the technical feasibility and usefulness of this novel approach to
pave the way for potential, future global change.

\subsubsection{IP address synchronization} 
IP synchronization between the two ends of an established session is necessary.  We use a pseudo-random number generator with a chosen seed to generate a different set of $N$ addresses.  To reduce the number of BGP route injections, we limit the generated addresses to be within a smaller range of randomly generated, or deliberately chosen ASes.  

The random number generator and seed are  and only a properly authenticated client can access via a covert reverse DNS channel~\cite{fu2016covert}.  The covert protocol allows us to transforms strings into legitimate DNS records.  The server encodes the message into a list of domain names and register them to a randomly chosen IP address.  The client does a reverse-DNS lookup on the IP address and decodes the domain names to retrieve the message.  Different from DNS tunneling, this does not use uncommon record types (e.g., TXT records) or carry suspiciously large volume of traffic as DNS payloads. On the contrary, the resulting traffic will be normal DNS lookup/reverse-lookup traffic, which will not attract attention. The data transmission is not vulnerable to DPI.



\section{Experimentation}\label{sec:exp}
A series of experiments were conducted to demonstrate the viability and networking capability of this innovative addressing paradigm.  The goal is to show that we can effectively maintain the communication session, while the end host adopts short-lived and randomized IP addresses.

\begin{figure*}[!htb]
\centering
\includegraphics[width=.9\textwidth]{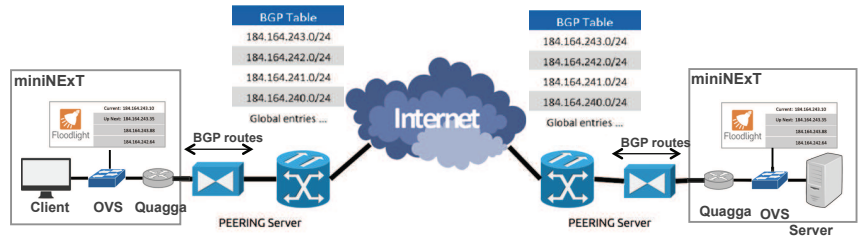}
\caption{Configuration of the experiment.}
\label{fig:con}
\end{figure*}

\subsection{Testbeds}
Our experiment is set up using the networking resources provided by GENI~\cite{geni} and the PEERING testbed~\cite{peering}.  GENI is a NSF-founded large-scale distributed testbed that gives academics and corporations access to various physical and virtual network resources.  With GENI, we can test communications between hosts in different geographic locations across the world.  The PEERING testbed is a real BGP testbed that connects to real networks to physical university networks and IXPs around the world.  Traditional inter-domain routing experiments were either based on passive observation of existing routes or on simulations, which greatly restricts the fidelity of the results.  PEERING removes these limitations by providing researchers a platform to conduct inter-domain routing tests in realistic environments.

\subsection{Experiment design}
The experiment consists of two parts: BGP route propagation and maintaining the communication session while end host's IP address changes.  The configuration of the experiment is shown in Figure~\ref{fig:con}, where the client and server sit on GENI racks located at NPS and NYSERNet, respectively.  On each rack, we installed a virtual network emulator called miniNExT~\cite{miniNExT}, which extends mininet~\cite{mininet} to easily build complex networks.  Each miniNExT includes a \textit{Floodlight manager}~\cite{floodlight} that rewrites packet headers, an \textit{OVS} that interacts with the Floodlight, and a \textit{Quagga router}~\cite{quagga} that runs a virtual BGP router to communicate with the external BGP routers of other ASes.

\subsubsection{BGP route propagation}
The PEERING testbed maintains multiple IP prefixes for experimental purposes.  Researchers can announce BGP prefixes at selected US and EU point of presences (PoPs) to the Internet.  As illustrated in Figure~\ref{fig:con}, the Quagga router at each end is attached to the upstream PEERING server via PEERING's OpenVPN tunnel.  PEERING runs its own Quagga software routers to establish a BGP session with their neighbors.  Instead of running the BGP route selection process, those PEERING servers provide PEERING clients with full control over BGP route announcements.  Our Quagga routers build a BGP peer session with PEERING's Quagga routers.  Any BGP announcement from our Quagga router can propagate across the Internet through the PEERING servers, and vice versa.  The PEERING server will just relay the announcements as long as the BGP announcement is confirmed valid by PEERING.  

Four IP prefixes were reserved for our experiment: 184.164.243.0/24 for server side and 184.164.242.0/24 for client side.  This provides two blocks of contiguous routable IP addresses.  

\subsubsection{Maintaining communication sessions}
In this experiment, only the server changes its address, and the client needs to keep track the server's IP address.  This conforms to the use case of censorship circumvention presented in section~\ref{sec:tarn}.  We leveraged a Floodlight OpenFlow controller that sits on each end to maintain the communication between client and server.  

Each Floodlight controller keeps an ordered list of IP addresses called \textit{external IP addresses} that the server is going to use within a certain time frame.  In our experiment, a fixed \textit{internal IP} is adopted, which the client/server process talks to directly.  This presents the illusion that the application is talking to the same IP all the time while the external IP changes.  Floodlight manages the flow rules that rewrite the packet header for port forwarding.  The Floodlight controller generates two types of packet: IP packets and ARP packets.  With each packet type, there are a pair of flows: one flow rewrites the outgoing packets by changing the internal IP address to the current external IP address; while the other flow rewrites the incoming packets and changes the external IP address to internal IP address.  In addition, Floodlight controller is also responsible for updating the external IP address at a pre-defined rate that is only known to the client.

\subsection{Experimental Results}
Since TARN is an ongoing project, only preliminary experimental results are provided.  It is assumed that a pseudorandom sequence of IP addresses are known to both the server and client.  It is also assumed that the server rapidly changes its IP address in a predetermined order that is known to both ends.   We have provided methods of IP synchronization between the client and server in section~\ref{subsec:challenge}.  The corresponding module will be implemented and integrated in future work.  The configuration of the experiment is consistent with the censorship circumvention use case in Figure~\ref{fig:tarn}, where the server's IP address changes at a fixed rate.

The objective of this experiment is to demonstrate two basic functions of TARN: (1) The SDN controller offers the capability to manipulate the hopping pattern of IP addresses, and (2) TARN maintains a stable and fluent session while the server frequently changes its IP address (111 different IP addresses from 184.164.243.0/24 with average dwell time between hops $<$ 10 sec).  


We had the client send out 672 IP packets and collected data received on the server.  
All the packets were successfully received by the client.  Given that the packets were sent at a stable rate, this means the dwell time for each IP address the server used were randomized.  
Moreover, any given distribution of IP hopping can be achieved using an SDN controller.  Although only preliminary results are available, they fully support that we can effectively maintain the communication session, while the end host adopts short-lived and randomized IP addresses. This lays the foundation for further experimentation, including hopping pattern manipulation and two-end IP hopping.



\section{Conclusion and Future Work}
\label{sec:con}
The TARN system is highly resistant DNS/IP filtering and traffic analysis, by removing a major vulnerability in today's addressing scheme -- the fixed binding between hosts and IP addresses, which has been used by cyber attacks to easily locate their victims.  TARN relies on the promising SDN technologies to realize stable and efficient routing between end hosts with dynamic addresses.  

We lay out the technical foundation and roadmap for a scalable implementation of this transformative paradigm.  Three different implementation strategies are provided, each of which performs IP hopping where appropriate to the users' specific requirements given the trust boundary.  We show that TARN is compatible with today's Internet infrastructure by properly addressing the possible challenges in deploying TARN.  

The one-way randomization version of the SDN-based solution was implemented and tested on PEERING's BGP testbed.  Since TARN is an ongoing project, only preliminary experimental results are provided in this paper.  The obtained results demonstrate that: (1) SDN controllers allow users to easily manipulate the IP hopping pattern to their own needs; and (2) it is completely feasible to maintain a stable and fluent communication between end hosts of dynamic IPs.  This has laid the foundation for further experimentation.

Future work includes (1) performing experiments to demonstrate the utility of the other designs, (2) completing the two-way IP randomization, (3) performing an in-depth security analysis of the SDN-based TARN solution, and (4) implementing TARN as the future of SDX.  We will perform extensive, invasive testing to rigorously prove that TARN is to IP/DNS filtering, DPI and side-channel attacks.  These results, in conjunction with our preliminary security analysis, show great strides towards a security-centered Internet without compromising performance.

%
%
%
%

\section*{Acknowledgment}
This material is based upon work sponsored by the National Science Foundation under Grant No. 1643020. Any opinions, findings, and conclusions or recommendations expressed in this material are those of the authors and do not necessarily reflect the views of the National Science Foundation.

\bibliographystyle{latex8}
\bibliography{latex8}

\end{document}